\documentclass[10pt, final,journal] {IEEEtran}

\usepackage{epsfig,cite}
\usepackage{setspace}
\usepackage{subfigure}

\usepackage{graphicx}
\usepackage{longtable}
\usepackage{amsmath}
\usepackage{color}
\usepackage[finnish,english]{babel}
\usepackage[T1]{fontenc}

\def\e{\begin{equation}}
\def\f{\end{equation}}
\def\%#1{\mbox{\boldmath $#1$}}
\def\=#1{\overline{\overline #1}}

\def\_#1{{\bf #1}}

\def\.{\cdot}

\def\##1{{\bf#1\mit}}

\def\am{\left(\begin{array}{c}}
\def\amm{\left(\begin{array}{cc}}
\def\a{\end{array}\right)}

\title{A Thin Electromagnetic Absorber for Wide Incidence Angles and Both Polarizations}

\author{Olli Luukkonen, Filippo Costa, \IEEEmembership{Student Member,~IEEE}, Agostino Monorchio, \IEEEmembership{Senior Member,~IEEE} and Sergei
A.~Tretyakov, \IEEEmembership{Fellow,~IEEE},
\thanks{O.~Luukkonen
and S.~A.~Tretyakov are with the Department of Radio Science and
Engineering/SMARAD CoE, TKK Helsinki University of Technology, P.O.
3000, FI-02015 TKK, Finland (email: olli.luukkonen@tkk.fi)}
\thanks{F.~Costa and A.~Monorchio are with the Department of Information Engineering,
University of Pisa, Via G.Caruso, 56122 Pisa, Italy}}

\begin{document}

\maketitle {\center \large }

\parskip 0pt

\begin{abstract}

In this paper a planar electromagnetic absorber is introduced whose
performance is maintained over a wide change of the incidence angle
for both TE and TM polarization. The absorber comprises an array of
patches over a grounded dielectric slab, with clear advantage in
terms of manufacturability. It is shown that a high value of the
relative permittivity of the substrate is essential for the
operation of the absorber. The main contribution of the paper is to
demonstrate and practically use the presence of an additional
resonance of high-impedance surfaces when the plasma frequency of
the wire medium comprising metallic vias in the dielectric substrate
is close to the original resonance of the high-impedance surface.
The presence of the vias between FSS and the ground plane is
discussed both for the case of a high-permittivity absorber and for
a low permittivity one. The radius of the vias influences the
oblique incidence TM absorption, and when properly designed, the
insertion of the vias result in bandwidth enlargement and higher
absorption.

\end{abstract}

\section{Introduction}

Classical structures for electromagnetic absorbers include Jaumann,
Salisbury, and Dällenbach absorbers \cite{Salisbury} (see also
\cite{Knott,Munk}). In Salisbury absorbers a resistive sheet is
placed at a distance of $\lambda/4$ over the ground plane in order
to generate losses to the incident field. In Jaumann absorbers a
resistive sheets are stacked over each other at an approximate
distance of a quarter wavelength (measured at the center frequency
of the absorption band) distance generating a wider absorption band
compared to the Salisbury absorber. In Dällenbach absorber the
structure is similar to the previous ones, with the exception that
no resistive sheets are used, but the incident power is dissipated
in lossy homogenous dielectric materials layered on top of each
other over a ground plane.

Possibilities to enhance the performance of these absorbers have
been widely studied. For instance, one may include chiral inclusions
to the Dällenbach absorber's dielectric coatings and use the chiral
resonance to enhance the absorption and enlarge the bandwidth of the
absorber \cite{Reinert}. One can also use complex fractal geometries
in a similar way \cite{Chandran}. In \cite{Terracher} a frequency
selective surface (FSS) was used on top of a grounded dielectric
substrate to widen the absorption band. In all of the aforementioned
absorption techniques, the widening of the absorption band is
achieved by creating an additional resonance in the vicinity of the
primary resonance, the $\lambda/4$-resonance of the grounded
dielectric slab. In these cases the thickness of the absorber
remains still considerably large.

Artificial impedance surfaces, or high-impedance surfaces, have been
used to create electrically thin electromagnetic absorbers. These
absorbers relate closely to the Salisbury absorber: the resonance is
achieved by using the properties of the high-impedance surface and
the absorption by a separate resistive sheet \cite{engheta}. The
resistive sheet can be realized by using commercially available
resistive materials on top of the capacitive sheet or between the
metallic parts of the capacitive sheet \cite{simms1,Mosallaei}, or
by connecting resistors between the adjacent metallic parts of the
capacitive sheet of the high-impedance surface \cite{gao,simms2}.
The drawback of these designs, especially the ones using lumped
resistors, is the inherent difficult way of realizing the resistive
sheet (i.e. the high cost of high frequency lumped resistors and the
number of spot welding). Slightly increasing the overall thickness
and exploiting two close resonances, a more wideband absorber can be
designed by these techniques \cite{Costa}.

In \cite{Tretyakov_motl} a simple way of realizing the rabsorbtion
behavior was introduced: one simply needs to add losses to the
grounded dielectric substrate. Further, in \cite{Tretyakov_motl} the
stability with respect to the TM-polarized incident angle was
obtained using metallic vias connecting the patches to the ground
plane. However, the analytical expressions in that paper for the
grid impedance of the array of square patches were not accurate. In
this paper we will show that with the revised expressions, angular
stability also for the TE-polarized incident fields is achieved with
sufficiently high values of relative permittivity of the substrate.
As a matter of fact, no vias are needed for electrically thin
substrates in order to realize angular stability. Furthermore, we
will show that the vias can be used to increase the absorption band
for the TM-polarized oblique incidence.

The rest of the paper is organized as follows: we study first
high-impedance surfaces without vias and show that the angular
stability of the absorbers is achieved by increasing the
permittivity of the substrate. We will then consider high-impedance
surfaces with vias and we will discuss the possibility to enlarge
the absorption band by using the effect due to the vias favorably.
For comparison we will discuss high-impedance surfaces with both
high- and low values of the relative permittivity.

\section{Surface impedance of the absorber without vias}

\begin{figure}[t!]
\centering \includegraphics[width=9cm]{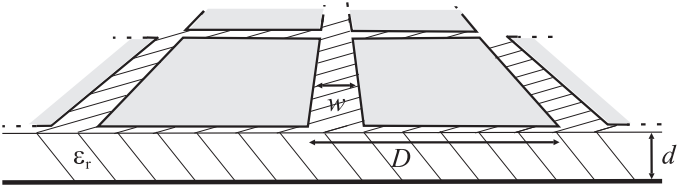} \caption{The
proposed absorbing structure. No metallic vias are embedded into the
substrate} \label{fig:1}
\end{figure}

The absorber structure is illustrated in Fig.~\ref{fig:1}(a). The
patch array over the grounded dielectric slab has a capacitive
response that, in conjunction with the inductive response of the
grounded dielectric slab, forms a resonant circuit. The patch array
comprises electrically small square metallic patches, so that the
structure is nearly isotropic. This type of artificial impedance
surfaces has been studied in our earlier work \cite{Luukkonen1}, in
which the surface impedance for the structure illustrated in
Fig.~\ref{fig:1} (a) was derived. The surface impedance of the
impedance surface, $Z_{\rm inp}$, can be considered to be a parallel
connection of the grid impedance of the patch array, $Z{\rm g}$ and
the surface impedance of the grounded dielectric slab, $Z_{\rm s}$:
\e Z_{\rm inp}^{-1} = Z_{\rm g}^{-1} + Z_{\rm s}^{-1}.\f For the
structure illustrated in Fig.~\ref{fig:1} the surface impedances for
TE and TM polarization read, respectively, \e Z_{\rm inp}^{\rm TE} =
\frac{j\omega\mu_0\frac{\tan\left(\beta d\right)}{\beta}}{1 -
2k_{\rm eff}\alpha \frac{\tan\left(\beta d\right)}{\beta} \left(1 -
\frac{\sin^2(\theta)}{\varepsilon_{\rm r} + 1} \right)},
\label{eq:Z_s^TE} \f \e Z_{\rm inp}^{\rm TM} = \frac{j\omega\mu_0
\frac{\tan\left(\beta d\right)}{\beta}\left(1 -
\frac{\sin^2(\theta)}{\varepsilon_{\rm r}} \right)}{1 - 2k_{\rm
eff}\alpha \frac{\tan\left(\beta d\right)}{\beta} \left(1 -
\frac{\sin^2(\theta)}{\varepsilon_{\rm r}} \right)}
\label{eq:Z_s^TM}, \f where $\beta=\sqrt{k_0^2\varepsilon_{\rm r} -
k_{\rm t}^2}$ is the normal component of the wave vector in the
substrate, $d$ is the height of the grounded dielectric substrate,
$k_{\rm eff} = k_0\sqrt{\varepsilon_{\rm eff}}$ is the wave vector
in the effective host medium (please see \cite{Luukkonen1} for more
details), $\varepsilon_{\rm eff} = \frac{\varepsilon_{\rm r} +
1}{2}$ is the effective permittivity of the host medium,
$\varepsilon_{\rm r}$ is the relative permittivity of the substrate,
and $\theta$ is the incident angle. Further, $\alpha$ is the {\it
grid parameter} for electrically dense ($k_{\rm eff}D\ll 2\pi$)
array of ideally conducting patches: \e \alpha = \frac{k_{\rm
eff}D}{\pi}\ln\left(\frac{1}{\sin\left(\frac{\pi w}{2D} \right)}
\right), \label{eq:alpha}\f where $D$ is the period of the structure
(see Fig.~\ref{fig:1}) and $w$ is the gap between the adjacent
patches. A more accurate approximation for the grid parameter can be
found in \cite{Yatsenko}. For electrically thin substrates we can
simplify \eqref{eq:Z_s^TE} and \eqref{eq:Z_s^TM} by using the
approximation $ \frac{\tan\left(\beta d\right)}{\beta} \approx d $.

We see that all angle-dependent terms in \eqref{eq:Z_s^TE} and
\eqref{eq:Z_s^TM} have the relative permittivity of the substrate
(see also \cite{Monorchio}), $\varepsilon_{\rm r}$, in the
denominator. This means that by increasing the permittivity of the
substrate we can diminish the effect of the incident angle to the
surface impedances. For relatively high values of $\varepsilon_{\rm
r}$ the expressions \eqref{eq:Z_s^TE} and \eqref{eq:Z_s^TM} for the
surface impedance both reduce in case of electrically thin
substrates to \e Z_{\rm s}^{\rm TE} = Z_{\rm s}^{\rm TM} \approx
\frac{j\omega\mu_0 d}{1 - 2k_{\rm eff}\alpha d},
\label{eq:approximation}\f which clearly is not a function of the
incident angle. By increasing the losses in the substrate (this
would affect the terms $k_{\rm eff}$ and $\alpha$), the
high-impedance surface structure could be used as an absorber that
has a stable operation with respect to the incidence angle.

The proposed high-impedance surface absorber is a resonant structure
and it suffers from the same characteristic features as other
resonators, that is from narrow bandwidth. If we write
\eqref{eq:approximation} in the lumped-element form, we have the
following expressions for the effective inductance and capacitance:
\e L_{\rm eff} = \mu_0 d, \label{eq:L_eff}\f \e C_{\rm eff} =
\varepsilon_0\left(\varepsilon_{\rm r} +
1\right)\frac{D}{\pi}\ln\left(\frac{1}{\sin\left(\frac{\pi
w}{2D}\right)}\right). \label{eq:C_eff}\f The losses of the surface
(due to the lossy substrate) are taken into account in the complex
value of the relative permittivity ($\varepsilon_{\rm r} =
\varepsilon_{\rm r}' - j\varepsilon_{\rm r}'')$. One can also
describe these losses with an effective resistor that would be
connected between the adjacent patches of the capacitive grid. The
conductance of the resistor can be calculated from the above
expression for the effective capacitance, and it reads: \e G_{\rm
eff} = \omega\varepsilon_0\varepsilon_{\rm
r}^{\prime\prime}\frac{D}{\pi}\ln\left(\frac{1}{\sin\left( \frac{\pi
w}{2D}\right)}\right). \f Furthermore, from the above expressions we
can see that, although the increase of relative permittivity
diminishes the dependence of the incident angle, this also increases
the effective capacitance and hence narrows the bandwidth. By
increasing the height of the substrate, the bandwidth of the
absorber can be somewhat increased, but increasing the height is not
desirable. Another possibility is to connect the patches to the
ground plane by metallic vias and increase the absorption band by
using the effect of these vias.

\section{Absorber with vias}

Let us now consider high-impedance surface absorbers in which the
metallic patches have been connected to the ground plane by vias. In
\cite{Tretyakov_motl} this case was considered in order to diminish
the angle-dependency from the absorber for TM-polarized incident
fields. However, in this paper we employ a different approach: we
use the high-permittivity substrate to diminish the angle-dependency
for both polarizations in the case of electrically thin slabs, and
use the vias to enhance the absorption and to widen the absorption
band.

If the metallic patches are not connected to the wires of the wire
medium layer, electric charges accumulate on the tips of the vias,
and the wire medium is spatially dispersive even for electrically
thin slabs \cite{Silveirinha}. For the absorber applications we wish
to suppress the spatial dispersion in the wire medium in order to
use it in the proposed design for an additional resonance. By
connecting large (compared to the via diameter) metallic patches to
the tips of the vias, we prevent the charges to be accumulated on
the tips of the vias. Instead, the charges spread over the metallic
patches. In addition, we need to have electrically thin slab of wire
medium so that the phase variation along the vias is minimum. With
these two conditions fulfilled, we can suppress and neglect the
spatial dispersion. In this case we model the wire medium slab as a
grounded uniaxial material slab \cite{Tretyakov, Luukkonen2} whose
normal component of the relative permittivity is calculated using
the local approximation (without spatial dispersion)
\cite{Tretyakov}: \e \varepsilon_{\rm n} = \varepsilon_{\rm
r}\left(1 - \frac{k_{\rm p}^2}{k_0^2\varepsilon_{\rm r}} \right),
\label{eq:epsn}\f where the plasma frequency can be calculated using
the quasi-static approximation \cite{Tretyakov}: \e k_{\rm p} =
\frac{1}{D\sqrt{\frac{1}{2\pi}\ln{\frac{D^2}{4r_0(D - r_0)}}}}. \f
Here $r_0$ is the radius of the vias. We can make use of the
artificial plasma resonance and widen the absorption band of the
high-impedance surface by choosing the plasma frequency of the wire
medium slab to lie close to the high-impedance surface resonance. In
the case of an electrically thin wire medium slab (except the very
proximity of the plasma frequency), the surface impedance of the
high-impedance surface reads for the TM polarization (see also
\cite{Luukkonen2}): \e Z_{\rm inp}^{\rm TM} = \frac{j\omega\mu_0
d\left(1 - \frac{\sin^2(\theta)}{\varepsilon_{\rm n}} \right)}{1 -
2k_{\rm eff}\alpha d \left(1 -
\frac{\sin^2(\theta)}{\varepsilon_{\rm n}} \right)}
\label{eq:Z_s^TMWM}. \f For the TE polarization we still have
\eqref{eq:Z_s^TE}.

Let us discuss the validity of \eqref{eq:Z_s^TMWM} in the absence of
losses. The local and quasi-static model of the wire medium assumes
that the phase variation along the normal direction of the uniaxial
slab is minimum. For an uniaxial slab, the normal component of the
propagation constant for the TM-polarized field reads (see
e.g.\cite{Tretyakov}): \e \beta_{\rm TM}^2 = k_0^2\varepsilon_{\rm
t}\mu_{\rm t} - k_{\rm t}^2\frac{\varepsilon_{\rm
t}}{\varepsilon_{\rm n}}, \label{eq:beta_TM^2}\f where
$\varepsilon_{\rm t}$ and $\mu_{\rm t}$ are the tangential
components of the relative permittivity and relative permeability,
respectively. In our case $\varepsilon_{\rm t} = \varepsilon_{\rm
r}$ and $\mu_{\rm t} = \mu_0$. We see that in the vicinity of the
plasma frequency of the wire medium ($\varepsilon_{\rm n}\rightarrow
0$), the normal component of the propagation constant
\eqref{eq:beta_TM^2} approaches infinity and invalidates our initial
assumptions on the quasi-static field distribution along the normal
direction within the uniaxial material slab. We see from
\eqref{eq:Z_s^TMWM} that in the case of relatively high values of
$\varepsilon_{\rm r}$ at low and high frequencies ($\varepsilon_{\rm
n} \rightarrow -\infty$ and $\varepsilon_{\rm n} \approx
\varepsilon_{\rm r}$, respectively) the surface impedance should
behave similarly to \eqref{eq:approximation}. However, very close to
the plasma frequency, the behavior of the surface impedance is very
different.

When losses are taken into account, we see that in the vicinity of
the plasma frequency ($\varepsilon_{\rm n} \rightarrow
-j\varepsilon_{\rm r}^{\prime\prime}$) the normal component of the
propagation constant \eqref{eq:beta_TM^2} does not tend to infinity
but to a certain complex value. By increasing the losses we can
clearly avoid the singularity of \eqref{eq:beta_TM^2} and therefore
widen the validity region of the approximation in the vicinity of
the plasma frequency. We cannot, however, determine the exact
boundaries of validity and it is considered to be outside of the
scope of this paper. We can hence conclude that \eqref{eq:Z_s^TMWM}
is valid for electrically thin substrates below and above a narrow
frequency band in the very vicinity of the plasma frequency (see
also \cite{Pamplona2008}).

\begin{figure}[t!]
\centering \subfigure[]{\includegraphics[width=9cm]{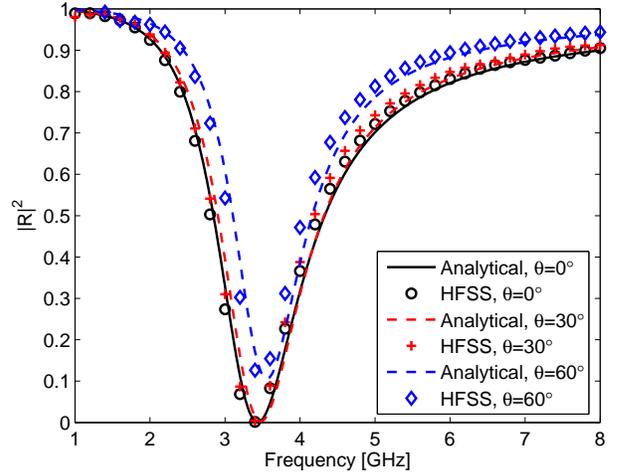} }
\subfigure[]{\includegraphics[width=9cm]{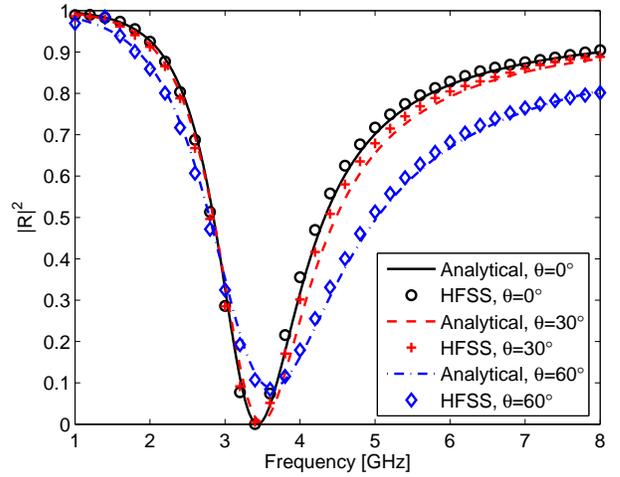}} \caption{The
power reflection factors for the incidence angles of
$0^\circ$,$30^\circ$, and $60^\circ$ for (a) TE and (b) TM
polarization. The parameters of the absorber are the following:
$D=5$\,mm, $w=0.1$\,mm, $h=3$\,mm, and $\varepsilon_{\rm r} = {\rm
9(1 - j0.222)}$.} \label{fig:2}
\end{figure}

\begin{figure}[t!]
\centering \subfigure[]{\includegraphics[width=9cm]{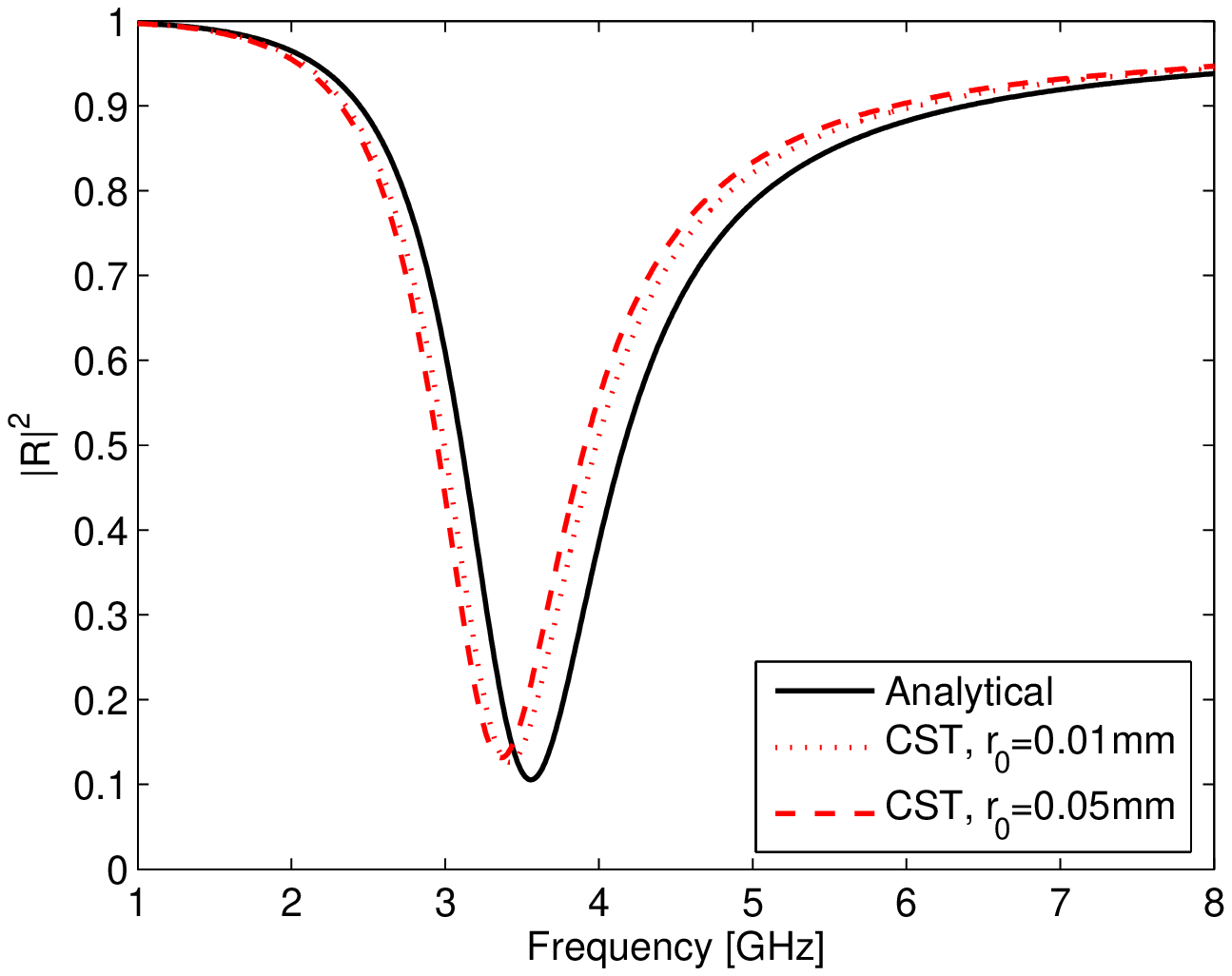} }
\subfigure[]{\includegraphics[width=9cm]{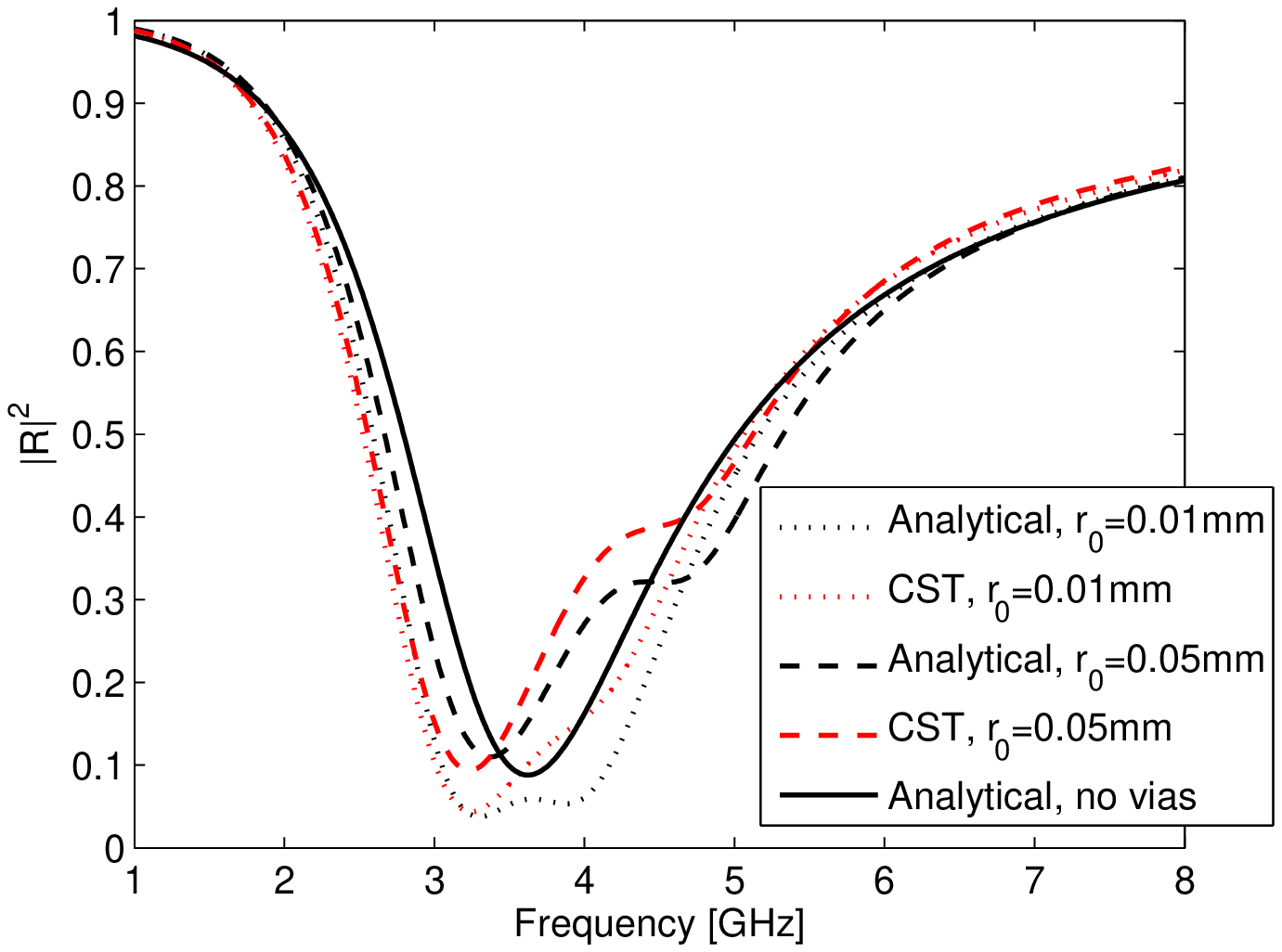}} \caption{The
effect of the vias to the power reflection factors for (a) TE and
(b) TM polarization. The angle of incidence is $60^\circ$. For the
TE-polarized case the analytical results are the same for different
via radiuses and in the absence of vias. $D=5$\,mm, $w=0.1$\,mm,
$h=3$\,mm, and $\varepsilon_{\rm r} = {\rm 9(1 - j0.222)}$.}
\label{fig:3}
\end{figure}

\section{Numerical results}

As an example of the performance of the absorbing layer, an
artificial impedance surface with the following parameters is
considered: $D=5$\,mm, $w=0.1$\,mm, $d=3$\,mm, and $\varepsilon_{\rm
r} = {\rm 9(1 - j0.222)}$. The power reflection factors are plotted
in Fig.~\ref{fig:2} for the normal incidence and for the angles of
$30^\circ$ and $60^\circ$ for both TE and TM polarizations. The
analytical results have been verified by full wave simulations done
using Ansoft's High Frequency Structure Simulator (HFSS)
\cite{ansoft}. The simulation results agree very well with our
analytical results and the performance of the absorber is little
affected by the change of the incidence angle, as expected. The
resonance frequency remains the same for both TE and TM
polarizations.

In the case of presence of vias, we wish to demonstrate the widening
of the absorption band by using the plasma resonance. The parameters
for the impedance surface are the same as in Fig.~\ref{fig:2}. The
radius of the vias, $r_0$, is changed in order to show that the
widening is truly because of the plasma resonance. In
Fig.~\ref{fig:3}(a) and (b) the power reflection factors are plotted
for the TE and TM polarization for the incident angle of $60^\circ$,
respectively. The following values were considered for the vias:
0.01\,mm and 0.05\,mm. The analytical results in
Figs.~\ref{fig:3}(a) and (b) have been verified using CST Microwave
Studio \cite{CST}.

\begin{figure}[h!]
\centering \subfigure[]{\includegraphics[width=9cm]{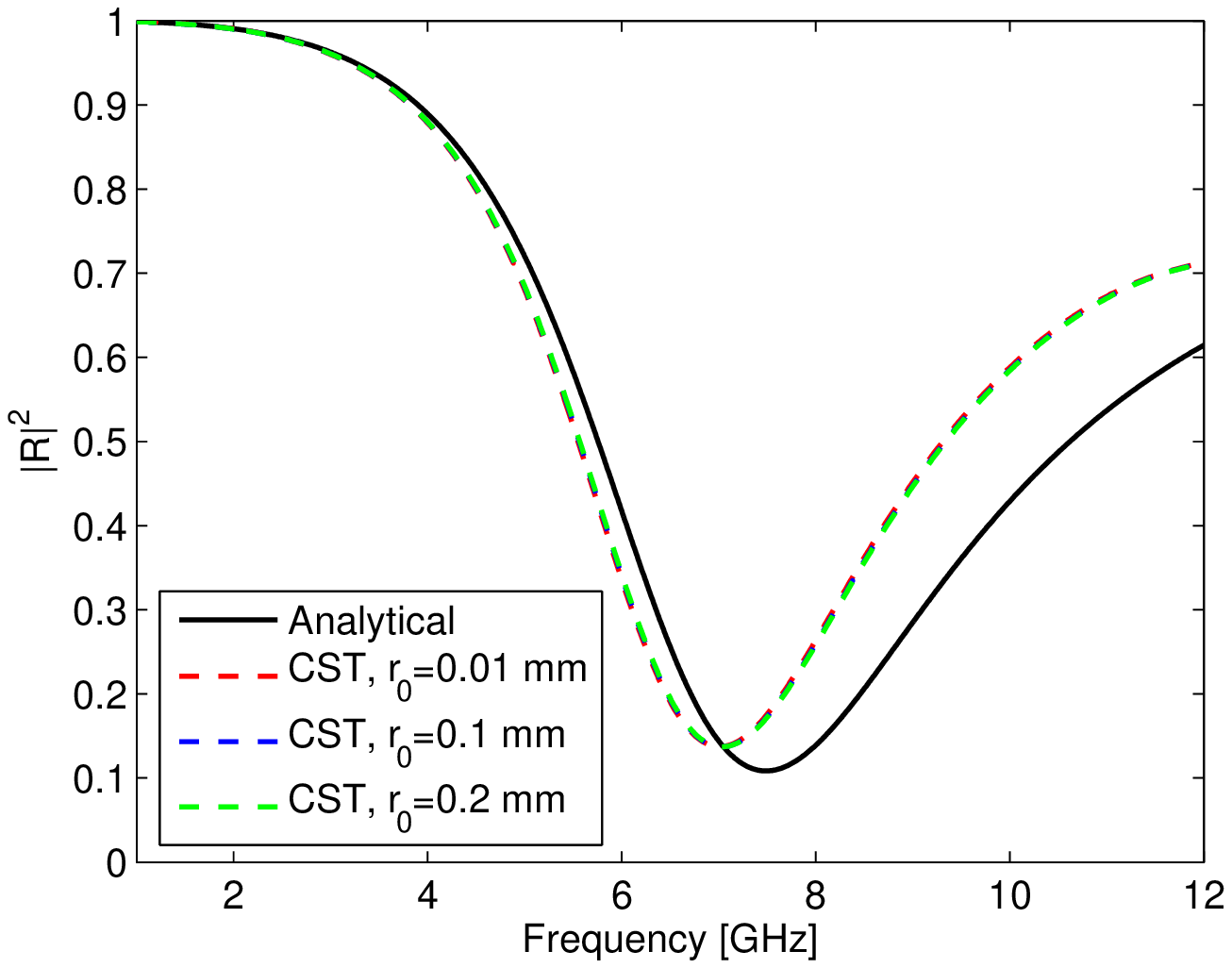} }
\subfigure[]{\includegraphics[width=9cm]{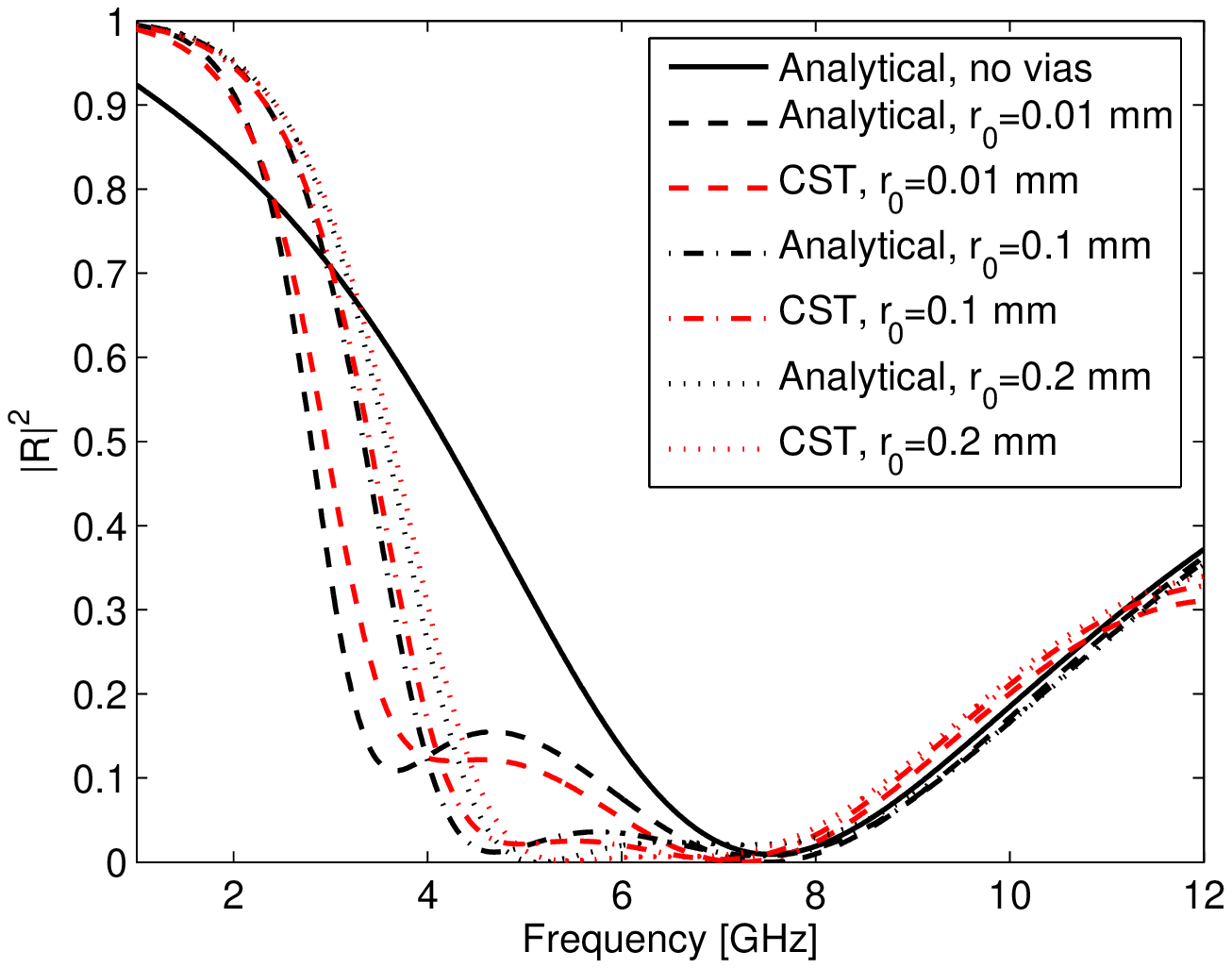}} \caption{The
effect of the vias to the power reflection factors for (a) TE and
(b) TM polarization. The angle of incidence is $60^\circ$. For the
TE-polarized case the analytical results are the same for different
via radiuses and in the absence of vias. The parameters of the
absorber are the following: $D=10$\,mm, $w=1.25$\,mm, $h=3$\,mm, and
$\varepsilon_{\rm r} = {\rm 2(1 - j0.5)}$.} \label{fig:4}
\end{figure}

\begin{figure}[h!]
\centering
\includegraphics[width=9cm]{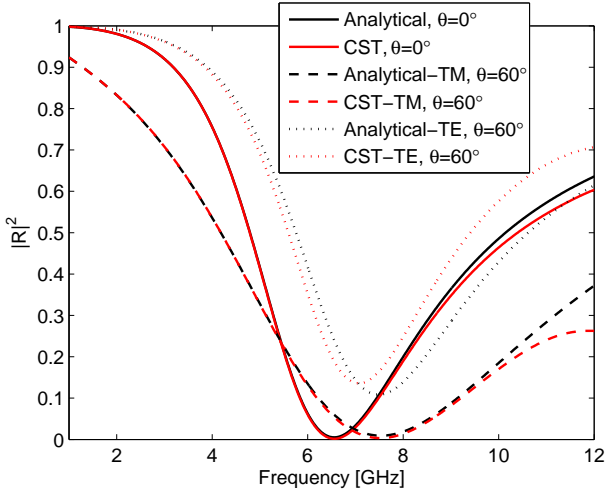}
\caption{The power reflection factors for the incidence angle of
$60^\circ$ for TE and TM polarization in the absence of vias. The
parameters of the absorber are the following: $D=10$\,mm,
$w=1.25$\,mm, $h=3$\,mm, and $\varepsilon_{\rm r} = {\rm 2(1 -
j0.5)}$.} \label{fig:5}
\end{figure}

\begin{figure}[h!]
\centering
\includegraphics[width=9cm]{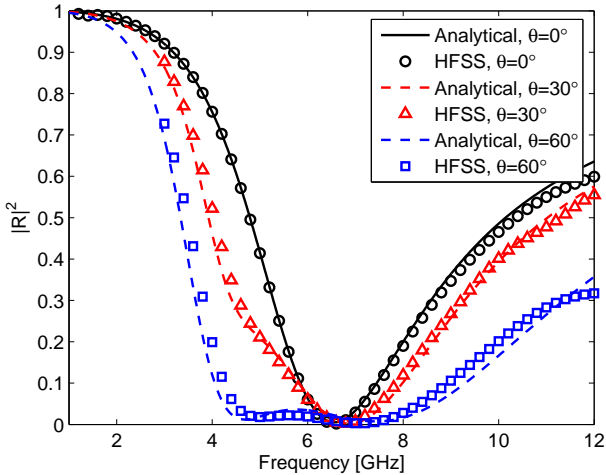}
\caption{The power reflection factors for different incidence angles
for the TM polarization in the presence of vias. The parameters of
the absorber are the following: $D=10$\,mm, $w=1.25$\,mm, $h=3$\,mm,
$\varepsilon_{\rm r} = {\rm 2(1 - j0.5)}$, and $r_0=0.1$\,mm.}
\label{fig:6}
\end{figure}

\begin{figure}[h!]
\centering
\includegraphics[width=9cm]{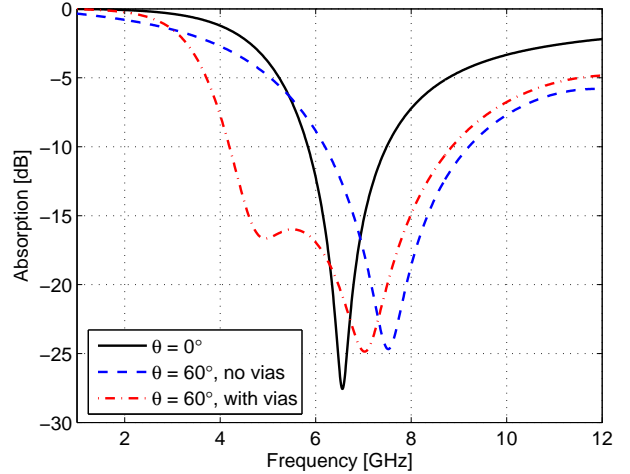}
\caption{The simulated absorption for different incidence angles in
decibels for the TM polarization in the presence and absence of
vias. The parameters of the absorber are the following: $D=10$\,mm,
$w=1.25$\,mm, $h=3$\,mm, and $\varepsilon_{\rm r} = {\rm 2(1 -
j0.5)}$. In the presence of vias $r_0=0.1$\,mm.} \label{fig:7}
\end{figure}

For comparison, a low-permittivity example is considered as well.
With this example we wish to demonstrate the dependency of the
operation on the relative permittivity of the substrate. In
addition, we wish to show that the bandwidth of the absorber can be
enlarged by means of the plasma frequency of the wire medium and by
choosing the parameters for the structure favorably. In
Fig.~\ref{fig:4}(a) and (b) the power reflection factors are plotted
for different via radiuses for an absorber with the following
parameters: $D=10$\,mm, $w=1.25$\,mm, $d=3$\,mm, $\varepsilon_{\rm
r} = {\rm 2(1 - j0.5)}$, and the incident angle $\theta=60^\circ$.
In Fig.~\ref{fig:4}(a) the difference between the simulation results
is minimum and the simulated results concur with each other almost
perfectly. It should be noted that here the substrates cannot be
treated as thin substrates nor is the patch array as electrically
dense as in the first example (Figs.~\ref{fig:2} and \ref{fig:3}).
Therefore, when calculating the surface impedances for the
absorbers, Eqs.~\eqref{eq:Z_s^TE}, \eqref{eq:Z_s^TM}, and
\eqref{eq:beta_TM^2} should be used. For the TE-polarized fields the
vias have no effect on the fields in the wire medium slab and the
normal component of the wave vector in the host medium can be used.
Fig.~\ref{fig:5} shows the results for the same absorber in the
absence of vias. Clear difference between the results for the
TM-polarized case is seen when Figs.~\ref{fig:4}(b) and \ref{fig:5}
are compared to each other. In Fig.~\ref{fig:6} the power reflection
factors are plotted for different incidence angles and in
\ref{fig:7} the simulation results obtained using CST Microwave
Studio for different incident angles are plotted in decibels.
Considering $-15$\,dB to be the limit of good absorption, we find
that in the case of vias for the oblique incidence of $60^\circ$ we
fulfil this limit in a frequency band from $4.6$\,GHz to 8\,GHz
whereas without vias this band would range from 6.8\,GHz to
8.3\,GHz.

In Table \ref{table1} the plasma frequencies for different via
radiuses is given for the considered structure. We can clearly see
the effect of the plasma frequency to the power reflection factors
in Figs.~\ref{fig:3} and \ref{fig:4}. We can also see that by
choosing the plasma frequency close to the resonance frequency of
the impedance surface, the absorption band can be enlarged and the
absorption can be enhanced.

In Figs.~\ref{fig:4} we see that the relative bandwidth of the
absorption has been increased when compared to the results presented
in Figs.~\ref{fig:3}. For the second example we have increased the
operational frequency of the absorber approximately by a factor of
two, lowered the permittivity of the substrate approximately by a
factor of four, and kept the height of the structure the same.
Because of this, the wave number for normal incidence in the
uniaxial material slab, \eqref{eq:beta_TM^2}, remains roughly the
same for both cases, as does the effective inductance
\eqref{eq:L_eff}. Simultaneously we have decreased the capacitance
of the structure.

The increase of the relative bandwidth can be partially explained
through the quality factor for a parallel resonant circuit. Although
the losses in our case are high, the following expression for the
bandwidth of parallel resonant circuits still holds qualitatively:
\e BW = Q^{-1} = G_{\rm eff}\sqrt{\frac{L_{\rm eff}}{C_{\rm eff}}} =
G_{\rm eff}\omega_{\rm r}L_{\rm eff}, \f where $\omega_{\rm r}$ is
the angular resonance frequency of the circuit. However, this is not
a fair comparison as the resonance frequencies of our example
"circuits" are not the same.

In our second example in Figs.~\ref{fig:4} the enlargement of the
bandwidth is also partially due to the fact that in the case of
lower permittivity substrates, the resonance frequency of the
absorber shifts to higher frequencies as the incident angle grows.
Together with the rather stable resonance caused by the stable
plasma frequency of the wire medium, this leads to the case where
the absorption band widens with the incidence angle, as shown in
Fig.~\ref{fig:6}.

\begin{table}[h!]
\centering \caption{The plasma frequencies for different via
radiuses.} \label{table1}
\begin{tabular}{c|c}
$r_0$ & $f_{\rm p}$ \\
\hline 0.01\,mm & 3.6\,GHz\\
\hline 0.05\,mm & 4.4\,GHz\\
\hline 0.1\,mm & 4.7\,GHz\\
\hline 0.2\,mm & 5.3\,GHz
\end{tabular}
\end{table}

%For the wide band absorber structure illustrated in Fig.~\ref{fig:2}
%the following parameters are used for the different designs:
%\begin{enumerate}
%\item $D=5$\,mm, $w_1=0.1$\,mm, $h=3$\,mm, and $\varepsilon_{\rm r} = {\rm
%9(1 - j0.222)}$.
%\item $D=5$\,mm, $w_2=0.5$\,mm, $h=3$\,mm, and $\varepsilon_{\rm r} = {\rm 9(1 -
%j0.222)}$.
%\end{enumerate} In Fig.~\ref{fig:4} the analytical
%results for the power reflection factors are plotted separately for
%the both designs. The simulated results are calculated for the
%complete structure.

%\begin{figure}[t!]
%\centering \includegraphics[width=8cm]{loss_WB.eps} \caption{The
%power reflection factor for the wide band absorber structure.}
%\label{fig:4}
%\end{figure}

\section{Conclusions}

An electrically thin absorber for wide incidence angles and for both
polarizations has been presented. The absorber is composed of a
patch array over a grounded dielectric substrate with or without
vias. It has been shown that a relatively high value of the
permittivity is needed for the substrate in order to have a stable
operation of the absorber with respect to the incidence angle. The
increase in the relative permittivity of the substrate leads to the
decrease in the bandwidth of the absorber. It has been shown in this
paper that the absorption band can be enlarged and the absorption
enhanced for the TM polarization by using metallic vias to connect
the metallic patches of the high-impedance surface to the ground
plane.

\section*{Acknowledgements}

Useful discussions with Profs. Igor Nefedov, Constantin Simovski,
M\'ario Silveirinha, and Alexander Yakovlev are warmly acknowledged.
The work was supported in part by the Academy of Finland and Tekes
through the Center-of-Excellence
Programme. %Olli Luukkonen wishes to thank the Jenny and Antti Wihuri
%Foundation, Emil Aaltonen Foundation, and Nokia Foundation for
%financial support.

\end{document}